\documentclass[reprint,
superscriptaddress,
amsmath,
amssymb,
prx
]{revtex4-2} 

\usepackage{graphicx}
\usepackage{physics}
\usepackage{color}
\usepackage{braket}
\usepackage{soul}
\usepackage[colorlinks=true, citecolor=blue, linkcolor=blue, urlcolor=blue]{hyperref}
\usepackage{comment}
\usepackage{xcolor}
\usepackage[normalem]{ulem}
\usepackage{xspace}

\usepackage{amsmath} 
\usepackage{bm}  

\newcommand{\kk}{\bm{k}}
\renewcommand{\qq}{\bm{q}}
\newcommand{\vq}{\nu\bm{q}}

\newcommand{\RR}{\bm{R}}
\newcommand{\vv}{\bm{v}}

\makeatletter

\begin{document}
\title{Data-driven compression of electron-phonon interactions} 

\author{Yao Luo}%
\affiliation{Department of Applied Physics and Materials Science, California Institute of Technology, Pasadena, California 91125, USA}

\author{Dhruv Desai}%
\affiliation{Department of Applied Physics and Materials Science, California Institute of Technology, Pasadena, California 91125, USA}

\author{Benjamin K. Chang}%
\affiliation{Department of Applied Physics and Materials Science, California Institute of Technology, Pasadena, California 91125, USA}

\author{Jinsoo Park}%
\affiliation{Department of Applied Physics and Materials Science, California Institute of Technology, Pasadena, California 91125, USA}

\author{Marco Bernardi}
\affiliation{Department of Applied Physics and Materials Science, California Institute of Technology, Pasadena, California 91125, USA}
\affiliation{Department of Physics, California Institute of Technology, Pasadena, California 91125, USA}
\email{bmarco@caltech.edu}

\begin{abstract}
\noindent 
First-principles calculations of electron interactions in materials have seen rapid progress in recent years, with electron-phonon ($e$-ph) interactions being a prime example.
However, these techniques use large matrices encoding the interactions on dense momentum grids, which reduces computational efficiency and obscures interpretability.  
For $e$-ph interactions, existing interpolation techniques leverage locality in real space, but the high dimensionality of the data remains a bottleneck to balance cost and accuracy. 
Here we show an efficient way to compress $e$-ph interactions based on singular value decomposition (SVD), a widely used matrix / image compression technique. 
Leveraging (un)constrained SVD methods, we accurately predict material properties related to $e$-ph interactions $-$ including charge \mbox{mobility,} spin relaxation times, band renormalization, and superconducting critical temperature $-$ while using only a small fraction (1$-$2\%) of the interaction data. 
\mbox{These findings unveil} the hidden low-dimensional nature of $e$-ph interactions. Furthermore, they accelerate state-of-the-art first-principles $e$-ph calculations by about two orders of magnitudes without sacrificing accuracy. 
Our Pareto-optimal parametrization of $e$-ph interactions can be readily generalized to electron-electron and electron-defect interactions, as well as to other couplings, advancing quantitative studies of condensed matter. 
\end{abstract}

\maketitle

%
%

\section{\label{sec1:intro}Introduction}
\vspace{-10pt}

Electrons in materials are subject to various interactions, including those with phonons, other electrons, and defects. Modeling of these interactions follows two main approaches $-$ analytic treatments that qualitatively capture the main physics with minimal models using only a few parameters, and first-principles calculations aiming at quantitative accuracy but often requiring specialized workflows, high computational cost, and large amounts of data.
A middle ground between these extremes would require formulating models of \mbox{electron} interactions that are economical, accurate, and interpretable. 
\mbox{Examples} of efficient models exist across domains $-$ in quantum chemistry, low-rank approximations~\cite{ERI2000,Pham2019,LU2015} 
can compress two-electron integrals to reduce the computational cost of post-Hartree Fock calculations~\cite{hohenstein2012,deprince2013} and extract the critical vibrational modes in a chemical reaction~\cite{pca-qmmm-1,pca-qmmm-2}; 
in correlated-electron physics, efficient parametrization of $e$-$e$ interactions~\cite{channelDcomp2009} enables the solution of functional renormalization-group flow~\cite{smfrg2012,smfrg2013} and the Bethe-Salpeter equation~\cite{Eckhardt2020,Wallerberger2021}. However, despite these isolated examples, it remains challenging to formulate widely applicable approaches to represent electron interactions both efficiently and accurately.
\\
\indent
Focusing on electron-phonon ($e$-ph) interactions, 
analytic treatments such as deformation potential for acoustic phonons~\cite{Bardeen1950,Herring1956} and the Fröhlich model for optical phonons~\cite{Fröhlich1954}, which use only a few parameters to describe $e$-ph interactions, are still widely utilized~\cite{Mahan2000,Ganose2021-nc}. 
In recent years, first-principles calculations of $e$-ph interactions using density functional theory (DFT)~\cite{martin_2004} and its linear-response variant, density functional perturbation theory (DFPT)~\cite{Giannozzi2001}, have enabled quantitative studies of properties ranging from transport to excited state dynamics to superconductivity~\cite{Bernardi2014,Li2015,Zhou2016,Liu2017,Jhalani2017,Ma2018,Zhou2018,Sohier2018,Zhou2019,Samuel2020,Lee2020_twophonon,Desai2021,Chang2022,Zheng2023}. 
%
Unlike the analytic models, in a typical first-principles calculation one represents the $e$-ph interactions using a multi-dimensional matrix with millions or billions of entries. 
This enormous number of parameters, which are computed rather than assumed, guarantees a faithful description of microscopic details such as the dependence on electronic states and phonon modes of $e$-ph interactions. Yet this complexity is also a barrier toward obtaining minimal models and tackling new physics. 
For example, materials with strong or correlated $e$-ph interactions need specialized treatments to capture polaron effects~\cite{Zhou2019,Lee2020_twophonon,yao-2022-LF,Giustino-2022-Polaron-PRL} and electron correlations~\cite{Zhou-2021-DFPT+U-PRL,Louie-2019-GWPT-PRL}. 
Reducing the high dimensionality of first-principles $e$-ph interactions would allow one to more efficiently describe this physics while retaining quantitative accuracy. 
The development of data-driven methods to tackle the high-dimensional Hilbert space in the many-electron problem, including neural network states~\cite{fermi-net,pauli-net} and tensor network methods~\cite{DMRG-schollwock-2005-RMP,DMRG-Garnet-2011,TNS-roman-2014,DMRG-Verstraete-2023-nature}, serve as inspiration. 
\\
\indent  
Here we show a low-rank approximation of first-principles $e$-ph interactions which significantly accelerates $e$-ph calculations while using only a small fraction (1$-$2\%) of the data and preserving quantitative accuracy. 
This is achieved by developing SVD calculations of $e$-ph matrices in Wannier basis to achieve a minimal representation of $e$-ph coupling.  
We use our compressed $e$-ph matrices to compute a range of properties, including charge transport, spin relaxation, band renormalization, and superconductivity, both in metals and semiconductors.  
Across all benchmarks, the highly compressed $e$-ph representation achieves a quantitative accuracy comparable to the standard workflow, while also providing a deeper understanding of the dominant patterns governing $e$-ph interactions. Principal component analysis sheds light on the inherent compressibility of $e$-ph coupling matrices.  
Recent interesting work on improving the efficiency of $e$-ph calculations~\cite{Anderson2020-prb,Deng2020-npj} is distinct in method and scope from our data-driven approach.
%

\section{results}
\subsection{compression of e-ph interactions}
\vspace{-7pt}
\indent
The key quantities in first-principles $e$-ph calculations are the $e$-ph matrix elements $g_{mn \nu}(\kk,\qq)$, which represent the probability amplitude for an electron in a band state $\ket{n\kk}$, with band index $n$ and crystal momentum $\kk$, to scatter into a final state $\ket{m \kk+\qq}$ by emitting or absorbing a phonon with mode index $\nu$, wave-vector $\qq$, energy $\hbar \omega_{\nu \qq}$, and polarization vector $\textbf{e}_{\nu \qq}$~\cite{Perturbo2021}:
\begin{equation}\label{gkq}
    g_{mn\nu}(\kk,\qq) =\sqrt{ \frac{\hbar}{2\omega_{\vq}}}\sum_{\kappa \alpha}\frac{\bf{e}^{\kappa \alpha}_{\vq}}{\sqrt{M_{\kappa}}} \braket{m \kk+\qq|\partial_{\qq \kappa \alpha }V|n\kk},
\end{equation}

\vspace{10pt}
\noindent
where $\partial_{\qq \kappa \alpha }V \equiv \sum_{p}e^{i\qq \textbf{R}_p}\partial_{p \kappa \alpha }V$ is the lattice-periodic $e$-ph perturbation potential, given by the change in the DFT Kohn-Sham potential with respect to the position of atom $\kappa$ (with mass $M_\kappa$ and located in unit cell at $\textbf{R}_p$) in the Cartesian direction $\alpha$. 
The inset in Fig. \ref{fig-1}(a) shows schematically such an $e$-ph scattering process. 
We separate the $e$-ph interactions into short- and long-ranged~\cite{Vogl,Mauri-2015-frohlich-PRB,verdi2015,Jhalani2020,jinsoo-quad-2020-PRB,Hautier-2020-Quadrupoles-PRL,Hautier-2020-Quadrupoles-PRB}, 
\begin{equation}
    g_{mn\nu} (\kk,\qq) = g^{L}_{mn\nu}(\kk,\qq) + g^{S}_{mn\nu}(\kk,\qq).
\end{equation}
The long-range part $g^{L}_{mn\nu}(\kk,\qq)$ includes dipole (Fr\"ohlich) and quadrupole contributions, which can be written analytically, using classical electromagnetism, in terms of Born effective charges and dynamical quadrupoles obtained from DFPT. 
The short-ranged part $g^{S}_{mn\nu}(\kk,\qq)$ cannot be written in closed form and needs numerical quantum mechanics to be computed, a consequence of the nearsightedness of electronic matter~\cite{Kohn_near}. Because $g^{S}_{mn\nu} (\kk,\qq)$ is a smooth function of electron and phonon momenta, it is short-ranged in a real-space representation using a localized basis set such as atomic orbitals~\cite{agapito} or Wannier functions~\cite{Giustino2007, Perturbo2021}. 
\\
\indent 
The short-range $e$-ph coupling matrix in Wannier basis, $g_{ij}^{\kappa\alpha}(\RR_e,\RR_p)$, is obtained by transforming DFPT results computed on a coarse momentum grid $(\kk_c,\qq_c)$~\cite{Perturbo2021}:
\begin{align}\label{grerp}
    g_{ij}^{\kappa\alpha}(\RR_e,\RR_p) = \frac{1}{N_{\kk_c}N_{\qq_c}}\sum_{mn\kk_c}\sum_{\qq_c}
    e^{-i(\kk_c\RR_e+\qq_c\RR_p)}\,  
    \\
    \times \ \mathcal{U}^\dagger_{im}(\kk_c+\qq_c) \Delta V^S_{mn,\kappa\alpha}(\kk_c,\qq_c)\,\mathcal{U}_{nj}(\kk_c),
    \nonumber 
\end{align}
where $\mathcal{U}$ is a unitary transformation from Bloch to Wannier basis, and $\Delta V^S_{mn,\kappa\alpha}(\kk_c,\qq_c) = \braket{m \kk+\qq|\partial_{\qq \kappa \alpha }V^S|n\kk}$ is the short-ranged part of the perturbation potential in Bloch basis.  
To separate acoustic and optical modes, we carry out a rotation in atomic basis:
\begin{equation}\label{seperation}
    g^{\mu \alpha}_{ij}(\RR_e,\RR_p) =  \sum_{\kappa} A^{\mu}_\kappa\, g_{ij}^{\kappa \alpha}(\RR_e,\RR_p),
\end{equation}
where $A^{\kappa}_\mu = \exp(i\frac{2\pi}{N_{\rm at}}\kappa \mu)$ adds a relative phase to different atoms in the unit cell, and $\mu \in (0,...,N_{\rm at}-1)$ labels phonon modes ($N_{\rm at}$ is the number of atoms in the unit cell).  
This way, $\mu=0$ corresponds to the acoustic subspace, where all the atoms in the unit cell move in phase, and  $\mu\neq 0$ labels the optical modes. Here and below, we use a collective index $F = (ij,\mu \alpha)$
to label Wannier orbital pairs $ij$ and phonon mode and direction $\mu \alpha$, simplifying the notation of the Wannier-basis $e$-ph matrices to $g^F(\RR_e,\RR_p)$. 
\\
\indent
When viewed as a matrix for each mode and orbital pair, $g^F(\RR_e,\RR_p)$ decays rapidly with lattice vectors $\RR_e$ and $\RR_p$ and has a typical size $N_{\RR_e}\times N_{\RR_p}$ ranging between $10^2 \times 10^2$ and $10^3 \times 10^3$.
After carrying out SVD on $ g^F(\RR_e,\RR_p)$, we obtain
\begin{equation}\label{full-svd}
    g^F(\RR_e,\RR_p) = \sum_{\gamma} s^F_\gamma u^F_\gamma(\RR_e)\, {v_\gamma^F}^{*}(\RR_p),
\end{equation}
where $s^F_\gamma$ is the singular value (SV) with index $\gamma$, and $u^F_\gamma(\RR_e) $ and ${v^F}^{*}_\gamma(\RR_p)$ are the left and right singular vectors, respectively. 
One can interpret $s^F_\gamma$ as the coupling strength between the generalized electron cloud $\sum_{\RR_e}\!u^F_\gamma(\RR_e) c^\dagger_{i}(\RR_e)c_{j}(\textbf{0})$ and phonon mode 
$\sum_{\RR_p}\!{v_\gamma^F}^{*}(\RR_p)\!\left(b^\dagger_{\mu \alpha}(\RR_p)+b_{\mu \alpha}(\RR_p)\right)$, where $(c^\dagger, c)$ are creation and annihilation operators for electrons and $(b^\dagger, b)$ for phonons. 
%
For each channel $F$, there is a total of $\min(N_{\RR_e},N_{\RR_p})$ SVs; we keep only the $N_c$ largest ones, resulting in a truncated, low-rank $e$-ph matrix $\tilde{g}$:
\begin{equation}\label{svd-grerp}
    \tilde{g}_{N_c}^F(\RR_e,\RR_p) = \sum_{\gamma=1}^{N_c} s^F_\gamma \, u^F_\gamma(\RR_e) \,{v_\gamma^F}^{*}(\RR_p).
\end{equation}
This matrix can be conveniently transformed to momentum space using 
\begin{align}
     \tilde{g}^{F}_{N_c}(\kk,\qq) &= \sum_{\RR_p,\RR_e}
e^{i\kk\RR_e+i\qq\RR_p}\,  \tilde{g}_{N_c}^F(\RR_e,\RR_p) \nonumber 
     \\
     & \approx \sum_{\gamma=1}^{N_c} s^F_\gamma \, u^F_\gamma(\kk)\, {v_\gamma^F}^{*}(\qq),
     \label{svd-gkq}
\end{align}
where $u^F_\gamma(\kk) \!=\! \sum_{\RR_e} e^{i\kk\RR_e} u^F_\gamma(\RR_e)$ and ${v_\gamma^F}^* (\qq) \!=\! \sum_{\RR_p} e^{i\qq\RR_p} v^F_\gamma(\RR_p)$ are singular vectors in momentum space. (Note that the long-range part of the $e$-ph matrix is added after interpolation of this short-ranged part.)
Eq.~(\ref{svd-gkq}) provides a generic parametrization of $e$-ph interactions, where by increasing the number of SVs one can systematically tune the accuracy and computational cost. 
According to the Eckart–Young-Mirsky theorem, the truncated matrix $\tilde{g}$ obtained from SVD is an optimal low-rank approximation of $e$-ph interactions, in the sense that it minimizes the Frobenius-norm distance between the original and low-rank $e$-ph matrices~\cite{brunton2022}. 
%
%
From a computational viewpoint, Eq.~(\ref{svd-gkq})  can greatly accelerate the calculation of e-ph interactions and the associated material properties, with a speed up by the inverse fraction of SVs kept in the truncated e-ph matrix. In most cases, we will keep only 1-2\% of SVs, resulting in a 50$-$100 times speed-up for the key step in e-ph calculations (see Appendix~\ref{app1:cca} for details).

\subsection{Error and Pareto-optimal interactions}
\vspace{-7pt}
To test the accuracy of the truncated $e$-ph matrix and its convergence with respect to the number of SVs ($N_c$), we define a relative error for the $e$-ph matrix averaged over electron bands and momenta, and phonon modes and wave-vectors: 
\begin{equation}\label{error-gkq}
    \epsilon_{g}(N_c) =\frac{\sum_{mn\nu,\kk\qq}|g_{mn\nu}(\kk,\qq)-\tilde{g}^{N_c}_{mn\nu}(\kk,\qq)|^2}{\sum_{mn\nu,\kk\qq}|g_{mn\nu}(\kk,\qq)|^2},
\end{equation}
where $\tilde{g}^{N_c}_{mn\nu}(\kk,\qq)$ is the low-rank (approximate) and
$g_{mn\nu}(\kk,\qq)$ is the full first-principles $e$-ph matrix. 
\begin{figure}
    \centering
    \includegraphics[width=1.0\linewidth]{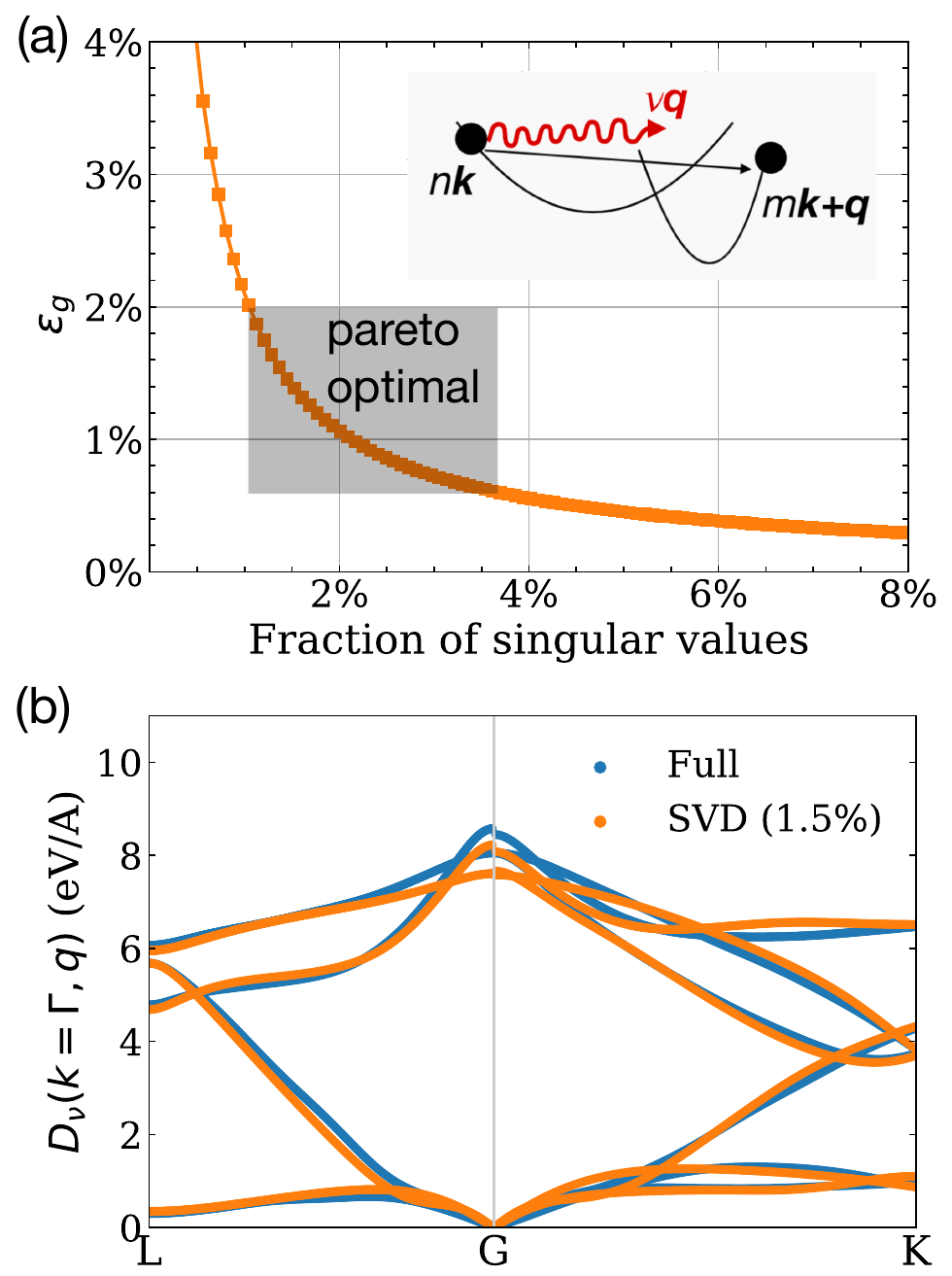}
    \caption{ (a) Error on the compressed $e$-ph matrix, computed using Eq.~(\ref{error-gkq}), as a function of the fraction of SVs used in the low-rank approximation. The Pareto-optimal region is shown with a shaded rectangle.
    (b) Mode-resolved $e$-ph coupling strength computed using the full $e$-ph matrix (blue) and the low-rank approximate matrix (orange) for silicon. The full $e$-ph matrix elements are computed on a real-space grid with size $N_{\RR_e} \!=\! 1325$ and $N_{\RR_p} \!=\! 1325$, the smallest values to achieve convergence, setting the electron momentum to $\kk\!=\!\Gamma$ and using the $N_b = 3$ highest valence bands. 
    In the SVD calculation, we keep 20 out of 1325 SVs, corresponding to a $\sim$1.5\% fraction of SVs, as noted in the legend.}
    \label{fig-1}
\end{figure}
Fig.~\ref{fig-1}(a) shows this error as a function of the fraction of SVs, $N_c/N_{\RR_p}$, kept in the approximate matrix. In the language of model selection~\cite{brunton2022}, the resulting curve of error versus number of parameters is the Pareto frontier for modeling $e$-ph interactions. We find that the error decreases rapidly with the number of SVs $-$ for example, $\epsilon_g$ is as low as $1\%$ when using only $2\%$ of SVs, which achieves a $98\%$ compression of the original $e$-ph matrix.  
This error curve defines a Pareto-optimal region, highlighted in Fig.~\ref{fig-1}(a), where $e$-ph calculations are both accurate and parsimonious~\cite{brunton2022}. This region spans 1$-$4\% of SVs in most of our calculations $-$ which corresponds to keeping $N_c \approx$ 10$-$50 SVs $-$ and suggests that many materials may possess only $\sim$10 dominant elementary $e$-ph interaction patterns. 
Accordingly, the $e$-ph coupling strength for each phonon mode~\cite{Sjakste2015,Perturbo2021},
\begin{equation}
\label{eq:Dnu}
D_{\nu}(\qq) = \hbar^{-1}\sqrt{2\omega_{\nu \qq}M_{\rm uc}\sum_{mn}|g_{mn\nu}(\kk=\Gamma,\qq)|^2/N_{b}}\,,
\end{equation}
(where $M_{\rm uc}$ is the mass of the unit cell and the band indices $m$ and $n$ run over $N_b$ bands), shown in Fig.~\ref{fig-1}(b), can be computed accurately using just the largest 1.5\% of SVs, matching closely results using the full $e$-ph matrix. 
\begin{figure*}
    \centering
    \includegraphics[width=1.02\linewidth]{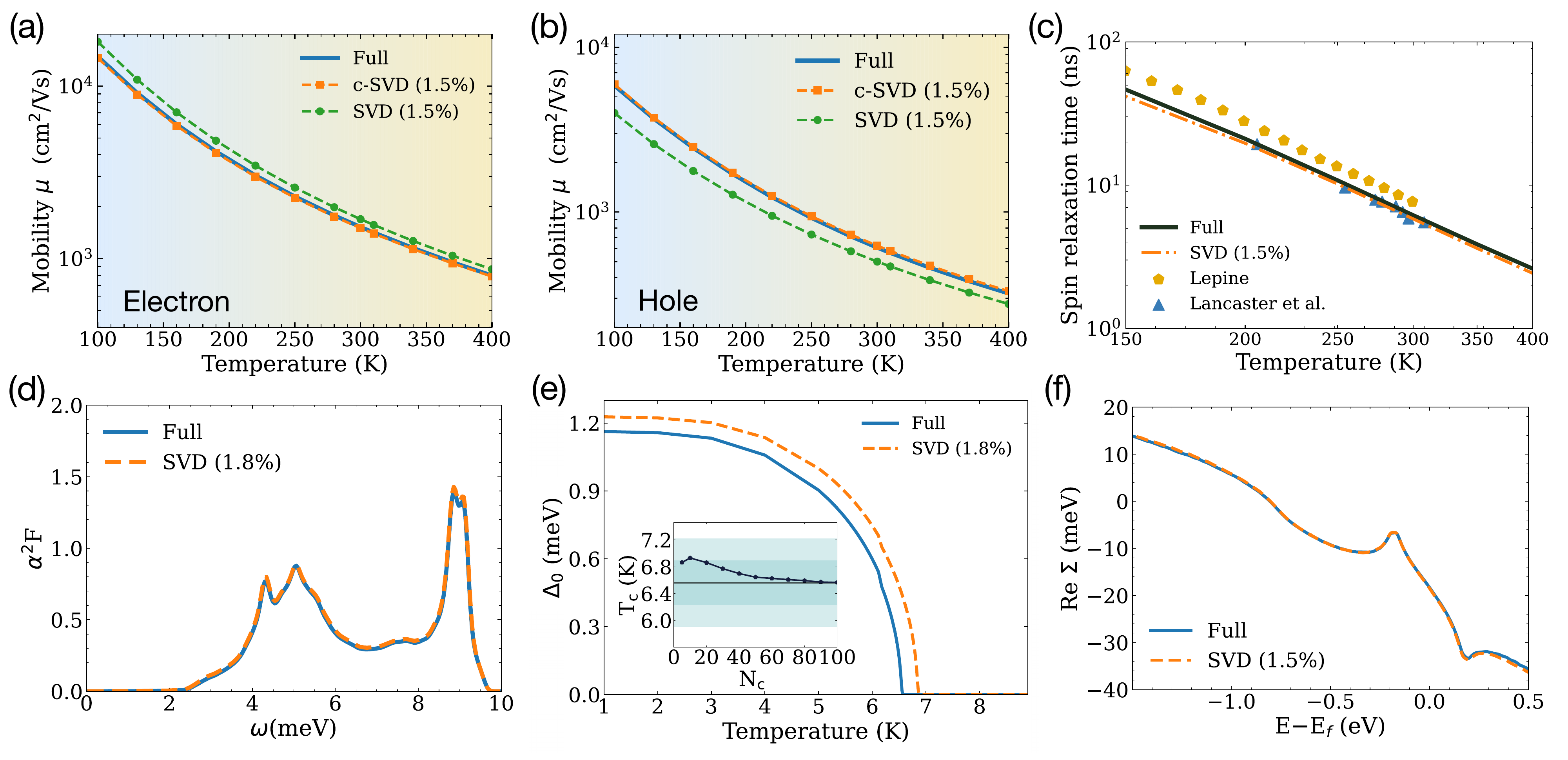}
\caption{ (a) Mobility of electrons in silicon, and (b) mobility of hole carriers in silicon, computed with the full $e$-ph matrix ($N_{\RR_e}\!=\!N_{\RR_p}\!=\!1325$) and compared with standard and constrained SVD, in both cases using 1.5\% of the SVs. 
(c) Spin relaxation times of electrons in silicon, computed with the full $e$-ph matrix ($N_{\RR_e}\!=\!N_{\RR_p}\!=\!1325$) and using SVD with 1.5\% of the SVs. Experimental data from Refs.~\cite{Lepine-srt-PRB-1970,Lancaster-srt-Proceedings-1964} are shown for comparison.\mbox{ 
(d) Eliashberg} spectral function $\alpha^2 F(\omega)$ for Pb, comparing full $e$-ph matrix  ($N_{\RR_e}\!=\!N_{\RR_p}\!=\!279$) with SVD results using 1.8\% of the SVs. (e) Superconducting gap $\Delta_0$ as a function of temperature, comparing full $e$-ph matrix results with SVD using 1.8\% of the SVs. The inset shows the convergence of the critical temperature $T_c$ with number of SVs;  the darker (lighter) colored regions indicate $5\%$ ($10\%$) error relative to the full calculation. 
(f) Band renormalization for electronic states near the Dirac cone in graphene at 20~K, comparing the full calculation with SVD using 1.5\% of the SVs.
}
    \label{Fig-2}
\end{figure*}
%
%
%

\subsection{Application to transport, spin and superconductivity}
\vspace{-7pt}
We showcase the accuracy of the low-rank approximate $e$-ph interactions by computing a wide range of material properties, including charge mobility, spin relaxation, phonon-assisted superconductivity, and phonon-induced band renormalization. 
Figs.~\ref{Fig-2}(a) and \ref{Fig-2}(b) show the electron and hole mobility in silicon for temperatures between 100$-$400~K, obtained using the full $e$-ph matrix and compared with SVD using 1.5\% of the SVs (see Appendix~\ref{app2:mobility}). The mobility is overestimated for electrons, and underestimated for hole carriers, despite the accuracy of the low-rank $e$-ph interactions in silicon (Fig.~\ref{fig-1}(a)). 
The error comes from the acoustic phonons, which interact weakly with electrons $-$ and therefore are ignored in the low-rank $e$-ph matrix $-$ but carry a considerable contribution to the mobility due to their large thermal occupation. 
To improve the treatment of acoustic phonons, we develop a constrained SVD (c-SVD) which preserves the deformation potential~\cite{acoustic-deformation-pot} for long-wavelength acoustic phonons in the compressed $e$-ph matrix (see Appendix~\ref{app3:csvd}). When using c-SVD, the mobility computed using only 1.5\% of the SVs is nearly identical to the full-matrix result for both carriers.  
\\
\indent
We also apply the low-rank approximation to spin-dependent $e$-ph matrices governing spin-flip $e$-ph interactions; these matrices enable first-principles calculations of spin relaxation times (SRTs) in centrosymmetric materials via the Elliot-Yafet mechanism~\cite{Jinsoo-EY-PRB-2020} (see Appendix~\ref{app4:srt}).   
The SRTs for electrons in silicon between 150$-$400 K are shown in Fig.~\ref{Fig-2}(c). %
Our results from SVD with $N_c\!=\!20$ (corresponding to $\sim$1.5\% of the SVs) match closely the full $e$-ph matrix calculations and agree with experimental results~\cite{Lepine-srt-PRB-1970, Lancaster-srt-Proceedings-1964}. 
Different from charge transport, standard SVD gives accurate SRTs in silicon because the optical phonons govern spin-flip processes. For materials where acoustic phonons contribute to spin relaxation, our c-SVD approach can be readily extended to the spin-dependent case.
\\
\begin{figure}
    \centering
    \includegraphics[width=1.0\linewidth]{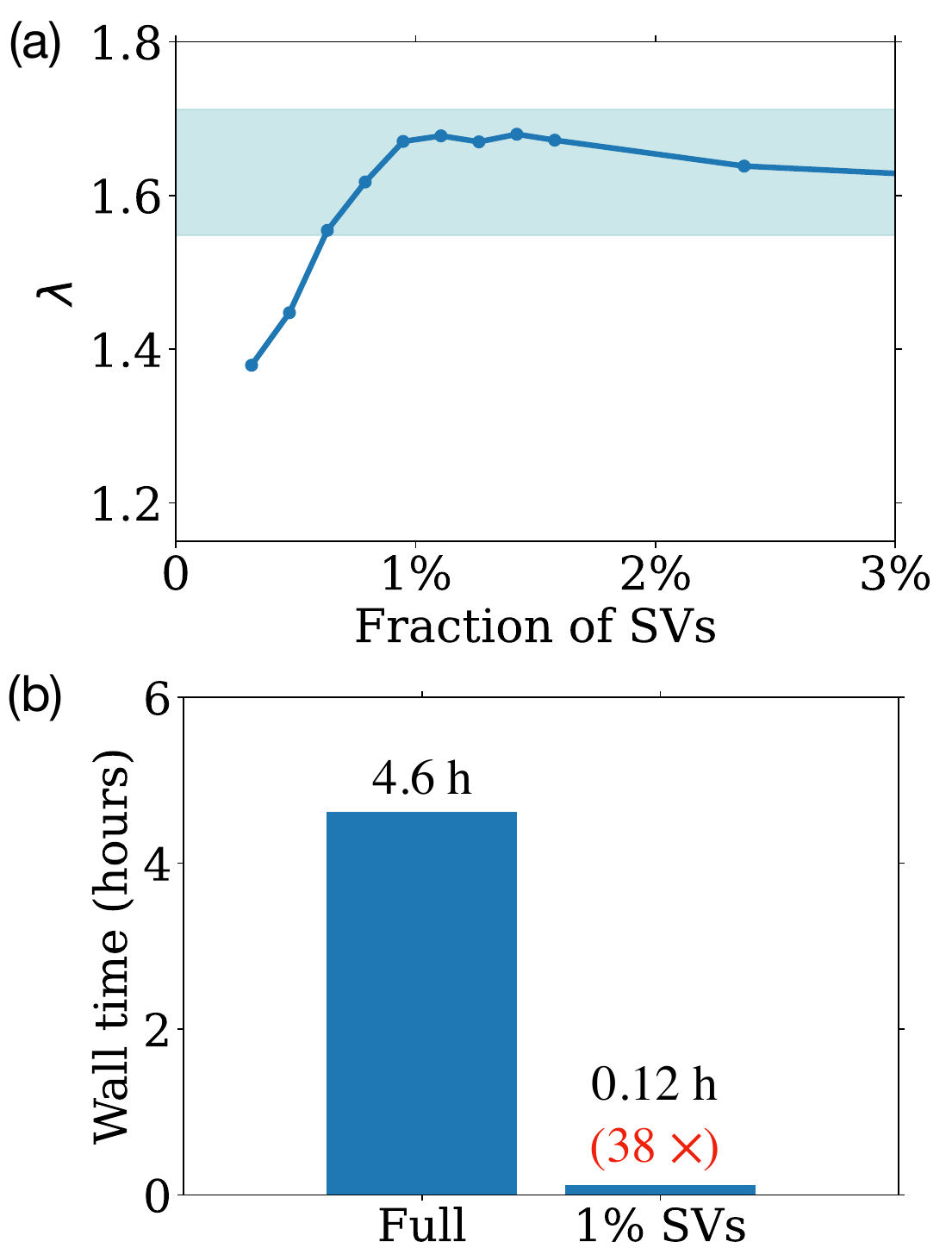}
    \caption{(a) $e$-ph coupling constant $\lambda$ in doped monolayer MoS$_2$ computed with our compression method using different fractions of SV. 
    The doping concentration is 0.22 electrons per formula unit. 
    The shaded region corresponds to an accuracy greater than 95\% relative to the fully converged calculation. 
    (b) Comparison of the wall time for computing the $e$-ph coupling constant $\lambda$ with the full $e$-ph matrices and with our SVD compression technique using 1\% of the SVs. 
    The 38$\times$ speed-up achieved by the SVD compression is indicated in red font.}
    \label{fig-mos2}
\end{figure}
\\
\begin{figure*}[ht]
    \centering
    \includegraphics[width=1.0\linewidth]{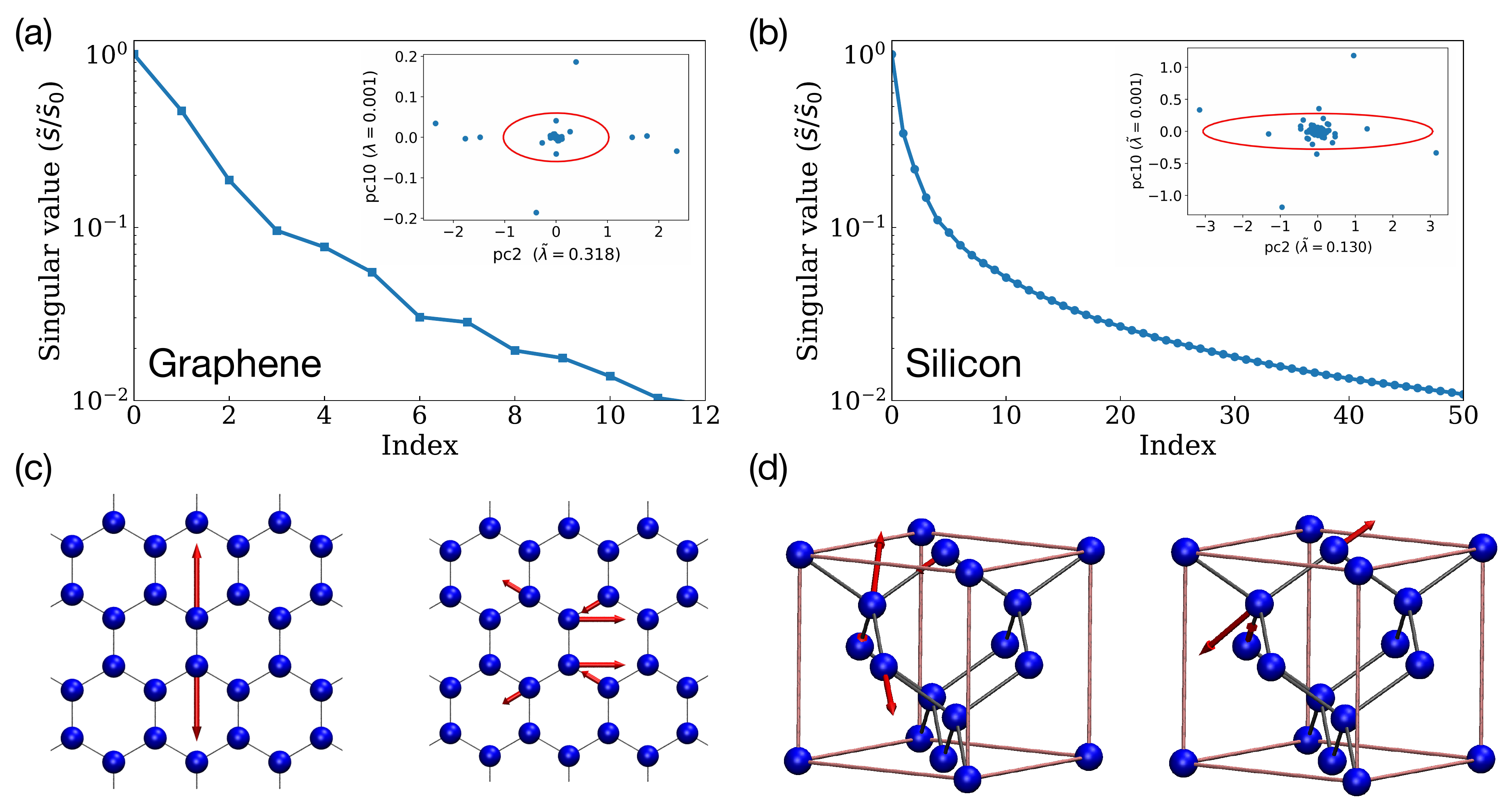}
    \caption{ 
    \mbox{(a) Decay of the SVs in graphene} and (b) decay of the SVs in silicon, shown by plotting the SVs referenced to the largest SV. Here, $\tilde{s}_i$ refers to the SVs averaged over Wannier orbitals and vibrational modes, $\tilde{s}_i = \sqrt{\sum_{F}(s^F_i)^2}$. 
    The respective insets show the real part of the 2nd versus 10th principal components, obtained from the PCA of the $e$-ph matrices; in parentheses we give the fraction of explained variance~\cite{mehta2019high}, $\tilde{\lambda}_i = \sigma^2_{i} / {\sum_{i}\sigma_i^2}$, where $\sigma_i^2$ is the variance of the $i$-th principal component. The red oval shows the standard deviation of the corresponding principal components, obtained by dropping feature vectors with norm smaller than $10^{-3} \times \tilde{s}_0$. 
    (c) Atomic vibrations associated with the dominant $e$-ph interactions in graphene, and (d) the same quantity in silicon, obtained by analyzing the phonon singular vectors, $\tilde{v}(\kappa\alpha\mathbf{R}_p)$ in Eq.~(\ref{eq:global-svd}), for the two largest SVs.\vspace{20pt}}
    \label{Fig-3} 
\end{figure*}
\indent
Calculations of phonon-mediated superconductivity, presented here using lead (Pb) as an example, can also leverage our low-rank approximation (see Appendix~\ref{app5:sc}).
Fig.~\ref{Fig-2}(d) compares the Eliashberg spectral function $\alpha^2F(\omega)$ from full $e$-ph matrix calculations with SVD results; using only the first $N_c\!=\!5$ SVs (here equal to 1.8\% of the SVs) suffices to reproduce the full calculation. 
We solve the isotropic Eliashberg equation self-consistently (see Methods for details) and compute the superconducting gap $\Delta_0$ as a function of temperature (Fig.~\ref{Fig-2}(e)) as well as the critical temperature $T_c$ versus number of SVs (inset of Fig.~\ref{Fig-2}(e)). 
The low-rank $e$-ph matrix with $N_c=5$ SVs provides a gap function in good agreement with the full-matrix calculation, which can be further improved by using a larger fraction of SVs; the critical temperature for $N_c=5$ SVs is very accurate, $T_c=6.9$ K which is within $5\%$ of the $T_c=6.6$ K value obtained using the full $e$-ph matrix. This result implies that as few as five elementary \mbox{$e$-ph interactions} determine $T_c$ in Pb. 
%
Finally, we compute band renormalization from e-ph interactions~\cite{grimvall-1981-ep} ] focusing on the contribution from the Fan-Migdal e-ph self-energy (see Appendix~\ref{app6:resigma}). The Debye-Waller term can also be added following Ref.~\cite{shang2023-jcp}. 
Results for graphene show that band renormalization near the Dirac cone can be computed keeping only the five largest SVs; this highly compressed $e$-ph matrix correctly predicts the kinks near the Dirac cone and matches full $e$-ph matrix results over a 2 eV energy range. 
We also carry out convergence tests with respect to the fraction of SVs included in the calculation for all the materials studied here (see the Supplemental Material~\cite{supplemental}). The rapid convergence with respect to the fraction of SVs guarantees that the full e-ph matrix calculation is not needed, and one can obtain accurate results by converging the desired property with respect to the fraction of SVs with minimal computational overhead. 
\\
\indent
While all the examples discussed above are for materials with nonpolar bonds, polar materials are even simpler to study with our compression method because the long-range (Frohlich) dipole contribution is dominant and is well modeled by an analytic formula using Born effective charges~\cite{Vogl,Mauri-2015-frohlich-PRB,verdi2015}. To illustrate this point, we demonstrate accurate mobility calculations in GaAs and PbTiO$_3$ using only 1\% of the SVs  (see the Supplemental Material~\cite{supplemental}).  
\indent

\subsection{Computational speed-up from compression} 
\vspace{-6pt}
We illustrate the computational speed-up achieved by our compression method using the $e$-ph coupling constant $\lambda$ in doped monolayer MoS$_2$ as a case study~\cite{doped-MoS2-prb-2013}. 
Following Ref.~\cite{doped-MoS2-prb-2013}, we employ a grid size of 288$^2$ $\kk$- and $\qq$-points for numerical integration and a Gaussain smearing of 0.002 Ry. We also 
leverage the improved Brillouin-zone sampling technique where only electronic states in a small energy window (0.006 Ry) near the Fermi surface are included in the calculation. 
In Fig.~\ref{fig-mos2}(a), we show the convergence of $\lambda$ with respect to the fraction of SVs. 
The shaded region corresponds to an accuracy greater than 95\% compared to the fully converged result. 
Similar to other quantities computed in this work, $\lambda$ converges rapidly with the fraction of SVs; in particular, using only 1\% of the SVs gives $\lambda$ within 2\% of the converged result. 
Fig.~\ref{fig-mos2}(b) compares the computational wall time for the calculation employing the full $e$-ph matrices and for our SVD compression method with 1\% SVs (note that both calculations use the same improved Brillouin zone sampling scheme). Our approach achieves a speed-up by 38 times relative to using the full $e$-ph matrices; if we count solely the time for $e$-ph interpolation, the speed-up is 83 times. 
This example illustrates the 1$-$2 orders of magnitude speed-up deriving from compressing the $e$-ph matrices with SVD. For a more detailed analysis of computational complexity, see Appendix~\ref{app1:cca}.

%
\vspace{-4pt}
\subsection{Dominant modes and\\ principal-component analysis}
\vspace{-10pt}
To understand the inherent compressibility of the $e$-ph matrices, we analyze the SV spectrum in graphene and silicon (Figs.~\ref{Fig-3}(a) and \ref{Fig-3}(b)). In both materials, the SVs decay rapidly, dropping by 1$-$2 orders of magnitude from the largest to the 10th largest SV. 
In principal component analysis (PCA)~\cite{brunton2022,mehta2019high}, this decay can be understood as a consequence of high-variance generalized directions in the $e$-ph matrix, $g^F(\mathbf{R}_e,\mathbf{R}_p)$, which capture the vast majority of the physics, while other principal components can be viewed as noise and neglected~\cite{mehta2019high}. 
We carry out PCA by treating each row of the matrix $g^F(\RR_e,\RR_p)$ as a feature vector, and find that the variance of the two leading principal components is one order of magnitude greater than for the 10th or following principal components, indicating that most of the physical information is already captured by the first few SVs (see the insets of Figs.~\ref{Fig-3}(a) and \ref{Fig-3}(b)). 
This analysis reveals that only a few atomic vibrational patterns dominate $e$-ph coupling. Although these dominant modes are not known a priori, they can be learned efficiently with SVD. 
We also apply the PCA to lead (see Fig. S4 in Supplemental Material ~\cite{supplemental}) ]) and observe a similar rapid decay of the SVs. We remark that the dimensionality reduction is general – it occurs in all the materials studied here, and it is associated to the rapid decay of the SVs, which we view as a consequence of the nearsightedness of electronic interactions.
\\
\indent
We visualize the atomic vibrations with dominant $e$-ph interactions by analyzing the vibrational singular vectors. To that end, we introduce a modified SVD that includes Wannier orbitals and phonon modes in the decomposition:  
\begin{equation}
\label{eq:global-svd}
    g_{ij}^{\kappa \alpha}(\RR_e,\RR_p) = \sum_\gamma s_\gamma\, \tilde{u}_\gamma(ij\RR_e)\,\tilde{v}_\gamma^*(\kappa\alpha\RR_p).
\end{equation}
In this global SVD, the singular vectors $\tilde{u}$ depend only on electron variables and $\tilde{v}$ only on phonon variables. 
This way, the phonon singular vectors $\tilde{v}^*_\gamma(\kappa\alpha\RR_p)$ can be interpreted as local vibrational modes (in the Wigner-Seitz cell associated with the coarse grid~\cite{Perturbo2021}) and visualized to study the dominant $e$-ph couplings. 
We show these singular vectors for the two modes with largest SVs in graphene and silicon in Figs.~\ref{Fig-3}(c) and \ref{Fig-3}(d), using arrows on each atom, with length proportional to the singular vector $\tilde{v}^*_\gamma(\kappa\alpha\RR_p)$, to indicate the atomic displacements in the modes obtained from SVD. 
\\
\indent
In graphene, where the electronic states consist of $p_z$ orbitals centered on each carbon atom, the dominant mode resembles a longitudinal optical phonon that brings the $p_z$ orbitals closer together in the unit cell.      
The second-strongest mode is a shear vibration resembling a transverse optical phonon, which spreads over multiple unit cells (Fig.~\ref{Fig-3}(c)). 
For the other modes, we observe that the vibrational pattern progressively delocalizes over multiple unit cells for decreasing values of the SVs. %
In silicon, where the electronic states consist of $sp^3$-like Wannier orbitals oriented along the chemical bond directions, the two modes with dominant $e$-ph coupling are associated with compression and stretching of the bonds (Fig.~\ref{Fig-3}(d)).
The intuition gained from this mode analysis can aid the formulation of model Hamiltonians in chemically and structurally complex materials, where keeping only the dominant $e$-ph interactions can provide effective models of transport and polaron physics informed by first-principles calculations~\cite{Nien-2020-Polaron-PRM,Yao-2022-Polaron-PRB,Giustino-2019-Polaron-PRL,van-de-walle-2014-prl,van-de-walle-2014-prb}. 
%

\section{Discussion and conclusion}
\vspace{-7pt}
\indent
The accuracy of the low-rank $e$-ph matrices implies that current brute-force first-principles calculations overparametrize $e$-ph interactions, falling too far on the right side of the Pareto-optimal region in Fig.~\ref{fig-1}(a). Conversely, textbook approaches such as Holstein and Fr\"ohlich models, which use only a handful of $e$-ph couplings, may fall short of achieving quantitative accuracy by using too few parameters. Our SVD compression in Wannier basis (followed by interpolation) provides a systematic route to achieve Pareto-optimal calculations. 
These optimal models enhance interpretability and enable a deeper understanding because they concentrate all the relevant $e$-ph physics in just a few parameters $-$ in our case, the leading SVs and singular vectors, which represent dominant $e$-ph interactions. 
%
%
\\
\indent
In summary, our results unveil the hidden low-dimensional nature of $e$-ph interactions. While accurate, current first-principles calculations overparametrize these interactions due to a lack of a priori knowledge of the dominant atomic vibrational patterns governing $e$-ph coupling. We have shown that when this optimal representation is achieved via SVD, using only 10$-$20 parameters (for each orbital pair and vibrational mode) is sufficient to obtain results with state-of-the-art accuracy.  
Surprisingly, this is only a small fraction (1$-$2\%) of the typical size of first-principles $e$-ph matrices. Compressing  $e$-ph interactions significantly accelerates calculations of material properties ranging from transport to spin relaxation to superconductivity. Future work will extend these ideas to other electronic interactions, with the goal of advancing \lq\lq precise but parsimonious\rq\rq~quantum many-body calculations in real materials.
Our approach works equally as well for small systems with a few atoms and for large systems with tens of atoms in the unit cell (see Fig.~S5 in Supplemental Material~\cite{supplemental}). In addition, as we plan to show elsewhere, our method enables many-body e-ph calculations that are currently inaccessible with standard Wannier interpolation, including the development of first-principles diagrammatic Monte Carlo, which sums e-ph diagrams to all orders from first principles, thus providing a gold standard for quantitative e-ph studies.

\begin{acknowledgments}
Y.L. thanks Ivan Maliyov and Junjie Yang for fruitful discussions. 
This work was supported by the National Science Foundation under Grant No. OAC-2209262. This research used resources of the National Energy Research Scientific Computing Center (NERSC), a U.S. Department of Energy Office of Science User Facility located at Lawrence Berkeley National Laboratory, operated under Contract No. DE-AC02-05CH11231.
\end{acknowledgments}

\appendix
\section{Computational complexity analysis }
\label{app1:cca}
\vspace{-6pt}
From a computational viewpoint, Eq.~(\ref{svd-gkq}) can greatly accelerate the calculation of $e$-ph interactions and the associated material properties. 
The key bottleneck in these calculations is obtaining the $e$-ph matrix elements on fine momentum grids, $g_{mn\nu}(\kk_f,\qq_f)$, starting from the Wannier representation, with a cost scaling as $\mathcal{O} (N_{\RR_p}N_{\kk_f} N_{\qq_f})$ for an optimal implementation~\cite{Perturbo2021} (for a fixed number of Wannier functions and atoms in the unit cell), where $N_{\kk_f}$ and $N_{\qq_f}$ are the number of points in the fine-momentum electron and phonon grids, with typical values of order $N_{\kk_f} \approx N_{\qq_f} \approx 10^6$.
In contrast, when using SVD, this interpolation step costs only $\mathcal{O} (N_c N_{\kk_f} N_{\qq_f})$, with a speed up by a factor $N_{\RR_p}/N_c$, the inverse fraction of SVs kept in the truncated $e$-ph matrix. In most cases, we will keep only 1$-$2\% of SVs, resulting in a 50$-$100 times speed-up for the key step in $e$-ph calculations. 
We show specific timing comparisons for all the materials studied in this work in Fig.~S1 of the Supplemental Material~\cite{supplemental}. In all cases, our algorithm achieves a speed up close to the ideal value of  $N_{\RR_p}/N_c$.
The memory improvement is also dramatic. A converged transport calculation in silicon requires a $\mathbf{k}$ grid of $100^3$ and $\mathbf{q}$ grid of $50^3$ points~\cite{Perturbo2021}; on these fine grids, the memory required to store the entire $e$-ph matrix $g_{mn\nu}(\kk_f,\qq_f)$ is 700 TB, while the memory needed to store the singular vectors $u_{\gamma}^F(\kk_f)$ and $v_{\gamma}^F(\qq_f)$ is only 128 GB when we retain $1.5\%$ of SVs, which guarantees accurate results as we show in Figs.~\ref{Fig-2}(a) and \ref{Fig-2}(b). 
This efficiency removes the key bottleneck in first-principles $e$-ph calculations.

\section{Mobility calculations}
\label{app2:mobility}
\vspace{-7pt}
The first-principles mobility calculations in silicon follows our previous work~\cite{jinsoo-quad-2020-PRB}. We include the quadrupole contribution analytically for silicon. The quadrupole tensor can be written as
\begin{equation}\label{quadruple}
    Q_{\mathrm{si},\alpha\beta\gamma} = (-1)^{\kappa+1}Q_{\mathrm{si}}|\epsilon_{\alpha\beta\gamma}|,
\end{equation}
where $\epsilon_{\alpha\beta\gamma}$ is the Levi-Civita tensor and the value of $Q_{\mathrm{si}} = 13.67$ is taken from Refs.~\cite{Royo2019-prx,GONZE2020107042}.

We compute the phonon-limited mobility at temperature $T$ using the BTE in the relaxation time approximation (RTA)~\cite{Perturbo2021}. 
We first obtain the $e$-ph scattering rate $\Gamma_{n\kk}$ using Fermi's golden rule, which is equivalent to using the imaginary part of the lowest-order $e$-ph self-energy~\cite{Bernardi2016}: 
\begin{align}
     \Gamma_{n\kk} &= \frac{2\pi}{\hbar} \frac{1}{\mathcal{N}_{\qq}} 
    \sum_{m\nu\qq} |g_{mn\nu}(\kk,\qq)|^2  \nonumber
    \\
     & \times [  (N_{\nu \qq}+1-f_{m \kk+\qq})
     \delta(\epsilon_{n\kk}-\epsilon_{m\kk+\qq}-\hbar \omega_{\nu \qq})
     \nonumber 
    \\
    & + (N_{\nu \qq}+f_{m \kk+\qq})\, \delta(\epsilon_{n\kk}-\epsilon_{m\kk+\qq}+\hbar \omega_{\nu \qq})], 
    \label{scattering}
\end{align}
where $\mathcal{N}_{\qq}$ is the number of $\qq$ points and $\delta$ is the Dirac delta function. 
Then we obtain the mobility from the BTE by summing over contributions from different electronic states and scattering processes~\cite{Perturbo2021}: 
\begin{equation}\label{mobility}
    \mu_{\alpha\beta}(T) = \frac{e}{n_c \Omega \mathcal{N}_{\kk}}
    \int dE\, \left(-\frac{\partial f}{\partial E}\right) \sum_{n\kk} 
    \tau_{n\kk}\, \vv_{n\kk}^\alpha \vv_{n\kk}^\beta \delta(E-\epsilon_{n\kk}),
\end{equation}
where  $\Omega$ is the volume of the unit cell, $\tau_{n\kk} = (\Gamma_{n\kk})^{-1}$ are relaxation times, $n_c$ is the carrier concentration, $f$ is the Fermi-Dirac distribution, and $\mathcal{N}_{\kk}$ is the number of $\kk$ points;
$\epsilon_{n\kk}$ and $\vv_{n\kk}$ are electron energies and band velocities, respectively. Our calculations in silicon  use a uniform grid with $200^3$ $\kk$ points and a uniform random grid with $10^5$ $\qq$ points, where $\kk$ and $\qq$ are electron and phonon momenta respectively. The delta function is approximated as a Gaussian  with a 10 meV smearing~\cite{Perturbo2021}.

\section{c-SVD algorithm }
\label{app3:csvd}
\vspace{-7pt}
Let us briefly describe our c-SVD algorithm. 
Similar to the acoustic sum rule (ASR) for the dynamical matrix~\cite{Giannozzi2001}, we formulate an ASR for the $e$-ph matrix elements: 
\begin{equation}
    g_{\nu_{A}}(\kk,\qq=0) = 0,
\end{equation}
where $\nu_{A}$ labels the acoustic modes, and we omit band indices for simplicity. 
The rationale for this $e$-ph ASR is that a rigid translation of the lattice will not change the electronic band structure. The real-space version of this $e$-ph ASR reads 
\begin{equation}
    \sum_{\kappa,\RR_p} g^{\kappa \alpha}_{ij}(\RR_e,\RR_p) =\sum_{\RR_p} g^{\mu=0, \alpha}_{ij}(\RR_e,\RR_p) = 0,
\end{equation}
where $g^{\mu=0, \alpha}_{ij}(\RR_e,\RR_p)$ accounts for the acoustic subspace of the $e$-ph matrix defined in Eq.~(\ref{seperation}). 
With this ASR, the $e$-ph matrix for long-wavelength acoustic phonons can be approximated to first order in $\qq$ as
\begin{align}
    \lim_{\qq \rightarrow0}\, g^{\mu=0,\alpha}(\RR_e,\qq) &= \lim_{\qq \rightarrow0}\, \sum_{\RR_p}g^{\mu=0, \alpha}(\RR_e,\RR_p) e^{i\qq\RR_p} 
    \nonumber 
    \\
    &\approx i\qq \cdot \textbf{A}^{\alpha}(\RR_e)\,,
\end{align}
where we defined a real-space deformation potential, which in general can be anisotropic, as 
\begin{equation}\label{deformation-pot}
   \textbf{A}^{\alpha}(\RR_e) \equiv \sum_{\RR_p} \RR_p g^{\mu=0, \alpha}(\RR_e,\RR_p).
\end{equation}

\noindent
In the limit of $|\qq| \to 0$, $\tilde{g}^{\mu=0,\alpha}(\RR_e,\qq)$ vanishes linearly in $|\qq|$, but the phonon occupation number diverges as $\frac{1}{|\qq|}$; therefore in the long-wavelength limit acoustic phonon scattering is often important. 
%
This acoustic phonon contribution is challenging to preserve when using the compressed $e$-ph matrices because standard SVD primarily captures large entries in the $e$-ph matrix. 
To address this point, we compress the $e$-ph matrix while conserving $\textbf{A}^{\alpha}(\RR_e)$ by imposing the following constraint:
\begin{equation}\label{constraint}
    \sum_{\RR_p} \RR_p\, \tilde{g}^{\mu=0, \alpha}_{N_c}(\RR_e,\RR_p) = \textbf{A}^{\alpha}(\RR_e).
\end{equation}
Satisfying this set of linear equations leads to a constrained low-rank approximation~\cite{CHU2003157}, a more general optimization problem.  
\\
\indent
The c-SVD is applied only to the acoustic subspace, which corresponds to $F=(ij,\mu=0\alpha)$, resulting in a compressed $e$-ph matrix of rank $N_c$:
\begin{align}\label{csvd-grerp}
    \tilde{g}^F_{N_c}(\RR_e,\RR_p) &= \tilde{g}^F_{N_c-3}(\RR_e,\RR_p) 
   \\
    &+ \sum_{\beta \beta' \in (x,y,z)} \left( \delta \textbf{A}^\alpha(\RR_e) \right)_\beta \lambda_{\beta\beta'} \left(\RR_p\right)_{\beta'},
    \nonumber 
\end{align}
where $\tilde{g}^F_{N_c-3}(\RR_e,\RR_p)$ is the truncated SVD of the $e$-ph matrix with $N_c-3$ singular values (see Eq.~(\ref{svd-grerp}));
$\delta \textbf{A}^\alpha$ is the residual term for $\tilde{g}^F_{N_c-3}(\RR_e,\RR_p)$, defined as 
\begin{equation}
 \delta \textbf{A}^\alpha (\RR_e)= \textbf{A}^\alpha(\RR_e) - \sum_{\RR_p} \RR_p\, \tilde{g}^{\mu=0, \alpha}_{N_c-3}(\RR_e,\RR_p),
\end{equation}
and $\lambda_{\beta\beta'}$ is the inverse of the overlap matrix between $\RR_p$ vectors:  
\begin{equation}
    \sum_{\beta'} \lambda_{\beta\beta'}\sum_{ \RR_p}\left(\RR_p\right)_{\beta'}\left(\RR_p\right)_{\beta''} = \delta_{\beta\beta''}.
\end{equation}
Using this approach, the compressed $e$-ph matrix $\tilde{g}^F_{N_c}(\RR_e,\RR_p)$ gives the same deformation potential as the full $e$-ph matrix in Eq.~(\ref{deformation-pot}), and its rank is smaller than or equal to $N_c$. This c-SVD workflow requires only a minimal computational overhead relative to standard SVD.

\section{Spin relaxation times}
\label{app4:srt}
\vspace{-7pt}
The first-principles calculation of SRTs in silicon follows our recent work~\cite{Jinsoo-EY-PRB-2020}. 
The spin-flip relaxation time $\tau^{\text{flip}}_{n\kk}$, for a band electron in state $\ket{n\kk}$, accounts for the Elliott-Yafet spin relaxation mechanism and is computed using~\cite{Elliott-1964,Yafet-1963,Jinsoo-EY-PRB-2020} 
\begin{align}
    \frac{1}{\tau^{\text{flip}}_{n\kk}} & = \frac{4\pi}{\hbar}\sum_{m\nu\qq}|g^{\text{flip}}_{mn\nu}(\kk,\qq)|^2 
     \nonumber 
    \\ 
    & 
    \times [  (N_{\nu \qq}+1-f_{m \kk+\qq})\,\delta(\epsilon_{n\kk}-\epsilon_{m\kk+\qq}-\hbar \omega_{\nu \qq})
    \nonumber
    \\
    &   
    + (N_{\nu \qq}+f_{m \kk+\qq})\, \delta(\epsilon_{n\kk}-\epsilon_{m\kk+\qq}+\hbar \omega_{\nu \qq})].
\end{align}
The key ingredients in this equation are the spin-flip $e$-ph matrix elements,
\begin{equation}
    g^{\text{flip}}_{mn\nu}(\kk,\qq) = \braket{m\kk+\qq\Downarrow|\Delta \hat{V}_{\nu\qq}|n\kk\Uparrow},
\end{equation}
 with $\Downarrow$ and $\Uparrow$ denoting nearly spin-down and nearly spin-up states, respectively. These matrix elements describe the probability amplitude to flip the spin of a band electron due to a particular phonon mode $\nu \qq$.
The macroscopic spin relaxation time, $\tau_s(T)$, is a thermal average over electronic states of the spin-flip scattering rates~\cite{Jinsoo-EY-PRB-2020}: 
\begin{equation}
    \tau_s(T) = \left\langle \frac{1}{\tau^{\text{flip}}_{n\kk}} \right\rangle^{-1}_T
     = \left( \frac{\sum_{n\kk} \frac{1}{\tau^{\text{flip}}_{n\kk}} \left(-\frac{df_{n\kk}}{dE}\right) }{\sum_{n\kk} \left(-\frac{df_{n\kk}}{dE}\right)} \right)^{-1}.
\end{equation}
The SRT calculations employ a uniform grid with up to $140^3$ $\kk$ points and a 5 meV Gaussian smearing for the delta functions. 

\section{Eliashberg spectral function and superconducting gap}
\label{app5:sc}
\vspace{-7pt}
We carry out DFT calculations on lead (Pb) using the generalized gradient approximation~\cite{Perdew1996} in the {\sc Quantum ESPRESSO} code~\cite{QE2009}. 
The ground state and electron wave functions are computed on a $14\times 14\times 14$ $\kk$-point grid with a kinetic energy cutoff of 90 Ry, and the lattice constant is set to 4.88~\AA. 
We use DFPT to calculate the phonon frequencies and eigenvectors, and the $e$-ph matrix elements $g_{mn\nu}(\kk,\qq)$, on coarse $6 \times 6 \times 6$ $\kk$- and $\qq$-point grids. 
We wannierize the 4 bands near the Fermi surface using the Wannier90 code~\cite{Marzari2014}, and obtain the $e$-ph matrices in Wannier basis using {\sc Perturbo}~\cite{Perturbo2021}. 
The Eliashberg spectral function is computed as
\begin{align}\label{a2f}
    \alpha^2F(\omega) &= \frac{1}{2}\sum_{\nu \qq} \omega_{\nu \qq} \, \lambda_{\nu \qq}\,\delta(\omega - \omega_{\nu \qq}),~~~~\text{with}    \nonumber 
    \\
\lambda_{\nu \qq} &= \frac{1}{N(\epsilon_{F})\omega_{\nu \qq}} \sum_{mn\kk} |g_{mn\nu}(\kk,\qq)|^2 
\\
&
~~~~~~~\times 
\delta(\epsilon_{n\kk} - \epsilon_F)\delta(\epsilon_{m\kk+\qq} - \epsilon_F), 
    \nonumber 
\end{align}
where $N(\epsilon_F)$ is the density of states at the Fermi energy ($\epsilon_F$). The Eliashberg function
$\alpha^2F(\omega)$ encodes the isotropic and retarded effective attraction between electronic states on the Fermi surface. 
Using $\alpha^2F(\omega)$, we obtain the gap function by solving the isotropic Migdal-Eliashberg Eq.~\cite{epw-2016}: 
\begin{align}\label{IMEE}
    Z(i\omega_j) &= 1 + \frac{\pi k_B T}{\omega_j}\sum_{j'} \frac{\omega_{j'}}{\sqrt{\omega_{j'}^2 + \Delta^2(i\omega_j)}}\lambda(\omega_j - \omega_{j'}), 
        \nonumber 
    \\
    Z(i\omega_j) &\Delta(i\omega_j) = \pi k_B T \sum_{j'} \frac{\Delta(i\omega_{j'})}{\sqrt{\omega_{j'}^2 + \Delta^2(i\omega_{j'})}} 
    \\
    &~~~~~~~~~~~~~~~~~~~~~~ \times \left[ \lambda(\omega_j - \omega_{j'}) - \mu_c^*\right], \nonumber 
\end{align}
where $T$ is the temperature, $\omega_j = (2j+1)\pi k_B T$ is the Matsubara frequency, $\mu_c^*$ is the screened Coulomb potential, $Z(i\omega_j)$ is the mass renormalization function, $\Delta(i\omega_j)$ is the superconducting gap function, and $\lambda(\omega_j) = \int_0^\infty d \omega \alpha^2 F(\omega)\frac{2\omega}{\omega_j^2+\omega^2}$ is the isotropic $e$-ph coupling strength.

For the numerical integrations in Eq.~(\ref{a2f}), we employ a $\kk$-point grid consisting of 400,000 quasi-random Sobol points (generated using {\sc SciPy}\cite{2020SciPy-NMeth}) and a $\qq$-point grid with 30,000 uniformly distributed random points;  
the delta functions are approximated as Gaussians with a $30$ meV smearing for electrons and $0.1$ meV smearing for phonons. 
Using the converged $\alpha^2 F(\omega)$ function, we set $\mu_c^*=0.1$ and solve Eq.~(\ref{IMEE}) iteratively for a range of temperatures. 
The critical temperature $T_c$ is obtained as the temperature where $\Delta_o = \Delta(i\omega_j=i\pi k_B T)$ extrapolates to zero. 

\vspace{30pt}
\section{Band structure \\ renormalization}
\label{app6:resigma}
\vspace{-7pt}
The DFT ground state calculation in graphene uses the local density approximation with a norm-conserving pseudopotential from Pseudo DOJO\cite{VANSETTEN2018}. We employ a 90 Ry plane-wave kinetic energy cutoff, a $60\times60\times1$ $\kk$-point grid, and a 2.46~\AA~lattice constant. 
For the DFPT calculation, we use coarse grids with $36 \times 36 \times 1$ $\kk$-points for electrons and $18\times18\times1$ $\qq$-points for phonons.   
The band structure renormalized by $e$-ph interactions, $\tilde{\epsilon}_{n\kk}$, is obtained as the DFT band structure plus the real part of the $e$-ph self-energy evaluated on-shell, 
\begin{equation}
    \tilde{\epsilon}_{n\kk} = \epsilon_{n\kk} + \Re \Sigma_{n\kk}(E = \epsilon_{n\kk}, T). 
\end{equation}
We use the lowest-order (Fan-Migdal) $e$-ph self-energy,
\begin{align}\label{realFM}
&\Sigma_{n\kk}(E,T) = \frac{1}{\mathcal{N}_{\qq}}\sum_{\nu\qq m} |g_{mn\nu}(\kk,\qq)|^2
\\
&\times \left[ \frac{N_{\nu\qq}+1-f_{m \kk+\qq}}{E - \epsilon_{m \kk+\qq} - \hbar \omega_{\nu \qq} - i\eta}
+ \frac{N_{\nu\qq}+f_{m \kk+\qq}}{E - \epsilon_{m \kk+\qq} + \hbar \omega_{\nu \qq} - i\eta}  \right], \nonumber 
\end{align}
where $\eta$ is a Lorentzian smearing~\cite{Park-2007-BandRenormalization-PRL}. 
We employ $10^7$ uniform random $\qq$-points for the numerical integration in Eq.~(\ref{realFM}) and set the Lorentzian smearing to 15 meV. 

For more accurate band renormalization calculations, one could use Wannier function perturbation theory (WFPT) to overcome errors resulting from finite number of Wannier functions~\cite{Lihm2021-prx}.

\providecommand{\noopsort}[1]{}\providecommand{\singleletter}[1]{#1}%

\end{document}